\documentclass[aps,prl,twocolumn]{revtex4}
\usepackage{graphicx}
\usepackage[usenames]{color}
\usepackage{epsfig}
\begin{document}
\baselineskip 11.74pt

\title{Chain Conformations and Photoluminescence in Poly(di-$n$-octylfluorene) }
\author{W. Chunwaschirasiri}
\author{B. Tanto}
\author{D.L. Huber}
\author{M.J. Winokur}
\affiliation{Department of Physics, University of Wisconsin, Madison, WI 53706}

\date{\today}

\begin{abstract}
The diverse steady-state spectroscopic properties of poly(di-$n$-octylfluorene) are addressed from a molecular-level perspective.  Modeling of representative oligomers support the experimental observation of at least three distinguishable classes of conformational isomers with differing chain torsion angles.  One class appears to be populated by a single compact structural isomer and this conforms to the so called $\beta$ phase.  A rigorous Franck-Condon analysis of the photoluminescence in conjunction with Frenkel-type exciton band structure calculations is performed.  These results accurately reproduce all major spectral features of the photoabsorption and those of singlet exciton emission.
\end{abstract}

\maketitle

Recent years have witnessed widespread activity in the field of polymer electronics.  Integral to this progress is development of realistic molecular-level models that can address fundamental issues of  photophysics and charge transport.  $\Pi$-conjugated organics, including small molecules, oligomers and polymers, that tend to pack into locally well-ordered crystalline structures have received broad attention because of notable advances\cite{Th:impact:admat01bredas,PPV:conform:prl02saxena,Th:PPVexciton:prl02molinari} in both theoretical chemistry and physics. These gains have not been matched by parallel progress in the burgeoning area of the functionalized polymers because of sample or processing derived heterogeneities and, more importantly, limited knowledge of the main chain and side chain conformations.   Thus many recent reports rely on qualitative underpinnings with few explicit references to the molecular level construction\cite{PFO:prompt:jpc01Bassler,PFO:alphabeta:cpl_Vardeny02,PFO:conjugation:prb04vardeny,PF:morph:prb04kohler}.  

Polyfluorenes (PFs)\cite{PFO:bluereview:macrorc01Neher} exemplify the extreme sensitivity of conjugated polymer photophyics to minor changes in molecular architecture and film-forming conditions.  Poly(di-$n$-octylfluorene), PF8 as sketched in the Fig.~1 inset, is best known for polymorphic phase behavior\cite{PFO:F8morph:prb00Bradley,PFO:processXRD:mm99Bradley} and the striking presence of a low energy emitting  ``$\beta$-type'' chromophore.  This often cited ``phase'' always appears as a minority constituent in the photoabsorption (PA) but dominates the optical emission\cite{PFO:F8morph:prb00Bradley,PFO:processXRD:mm99Bradley}.  Its photoluminescence (PL), by polymer standards, is also highly irregular in that these features become remarkably sharp at reduced temperatures\cite{xxpfo:trexciton:prb03bradley,pfo:pf8band:prb03mjw,PF:morph:prb04kohler}. 
Cadby et al.\cite{PFO:F8morph:prb00Bradley} have proposed that this emission originates
in regions of enhanced chain planarity. Unusual temperature-dependent properties and other emitting chromophores appear as well\cite{pfo:efficiency:jpcm02bradley,pfo:pf8band:prb03mjw,PFO:aging:mm03winokur}.
 
This letter advances a near molecular-level framework for explaining the diverse steady-state optical properties of singlet excitons in PF8. A simple empirical force-field modeling of representative oligomers yields a suggestive trifurcation of the low energy chain structures to give three nominally distinct classes or families of conformational isomers (referred to as $C_\alpha$, $C_\beta$ and $C_{\gamma}$) with 10-20$^\circ$ steps in the average fluorene-fluorene torsion angles ($|\phi|$).  The $C_\beta$ type ``family'' is comprised of just a single isomer.  The inferred nature of these families is then used to frame Frenkel-type tight-binding (t.b.) band structure calculations of the single chain PA and PL spectra.   Calculated optical spectra, including a full Franck-Condon (FC) vibronic progression, manifest many key attributes seen in the data. As a general feature PF8 $\beta$-type ($C_\beta$) emission incorporates a power law lineshape and not the composite Gaussian/Lorentzian typically assumed\cite{MEHPPV:PLfit:prb91hagler,PF:morph:prb04kohler}.  This approach also provides a  clear microscopic starting point for more quantitative analysis.

All of the PF8 polymers in this study were obtained from American Dye Source and used as received.  Various solutions were prepared, typically 1\% w/w in solvent, and the specific formulations (and some optical spectra) are already published\cite{pfo:pf8band:prb03mjw,PFO:aging:mm03winokur}.   Thin-film samples were cast onto sapphire substrates.  All PA and PL spectra were recorded from the same physical location under steady-state conditions using a dual PA/PL  spectrometer\cite{pfo:pf8band:prb03mjw}.  Thus all PL spectra are corrected for self-absorption and, additionally, normalized for detector response.    

\begin{figure} [b]
\begin{center}
\includegraphics[width=3.15in]{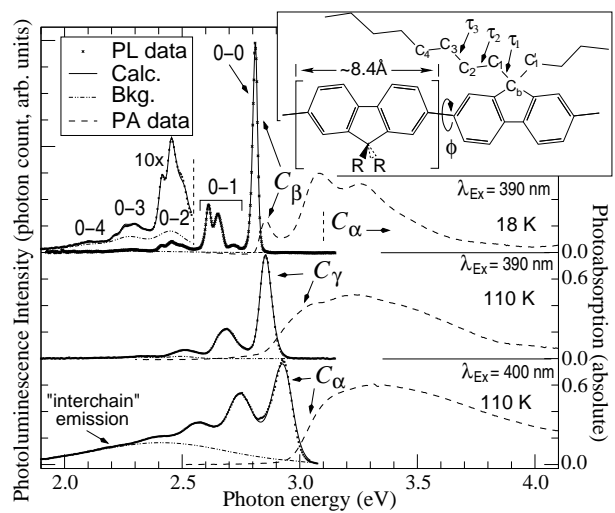}
\caption{Representative PF8 PA and PL spectra from the three claimed conformational
isomer families in combination with fits to the data using Eq.~1.  Inset: Sketch of PF8.}
\end{center}
\end{figure}
The three sets of PL/PA spectra in Fig.~1 demonstrate the wide range of PF8's optical properties.  Each PL spectrum has its emission dominated by one of the three claimed conformational isomer families.  The most intense ``0-0" peak is a superposition of the $\pi$-$\pi^*$ transition and, on the low-energy tailing edge, vibronic overtones from low energy torsional modes\cite{OPV:subband:jcp02Gierschner}.  Additional peaks arise from other FC vibronic subbands. The top spectra are representative of PF8 films\cite{PFO:aging:mm03winokur} spin cast from $p$-xylene and contain a coexistence of $C_\alpha$ and $C_\beta$ isomers.  Because of large scale exciton energy migration the lowest energy chromophores dominate the PL (here $C_\beta$ isomers).  The middle spectra are from a $C_{\gamma}$ isomer dominated sample and, in this case, the $\pi$-$\pi^*$ transition energies are intermediate to those of the proposed $C_\alpha$ and $C_\beta$ families.  This occurs in samples cycled above $\sim$430 K\cite{PFO:aging:mm03winokur}.   The bottommost spectra, obtained from casting PF8 dissolved in a toluene/tetrahydrofuran mixture\cite{pfo:pf8band:prb03mjw}, is dominated by emission from $C_\alpha$ type isomers.  This nematic glass sample also includes a broad low-energy  PL background due to interchain excitations\cite{PFO:PLdefect:apl02meijer}. Formation of the $C_\beta$ conformer has been suppressed by quenching to -30 $^\circ$C from a thermotropic liquid crystal state.  All PL spectra are notably sharper than the respective PA (again due to energy migration) but the distinct narrowing of the $C_\beta$ emission is quite evident.

Also included in Fig.~1 are non-linear least-square fits to the PL data using a slightly modified FC formula:  
\begin{eqnarray} PL(\hbar \omega) \propto (n(\omega)\hbar \omega)^3 \sum_{n_1=0}^{\infty} \cdots \sum_{n_p=0}^{\infty} \prod_{k=1}^{p} (1+c) \hspace*{0.35in} \nonumber \\  \left[\frac{e^{-S_k} S_k^{n_k}}{n_k!}\right]  \Gamma\ ( \delta[\hbar \omega  - (\hbar \omega_o - \sum_{k=1}^{p} n_k \hbar \omega_k)]) \end{eqnarray}
where $n(\omega)$ is the index of refraction (with data taken from  Ref.~\onlinecite{PF:morph:prb04kohler}), $\hbar\omega_0$ is the $\pi$-$\pi*$ transition energy, $\hbar\omega_k$ are vibrational mode energies for each mode $k$ with $n_k=0,1,2,...$ vibronic overtones.  $S_k$ is the conventional Huang-Rhys (HR) coefficient but $c=0.2(p-1)$ (for $p\ge 1$) is an ad-hoc term that increases the relative contribution by higher order modes.   Because the 18 K $C_\beta$ type PL is exceptionally well-defined nine separate base $\omega_k$'s were necessary (and extending up to seven overtones). When possible these frequencies were derived from Raman scattering data\cite{ltraman}.   The lineshape operator, $\Gamma$, is actually a composite function with, for example in the 0-0 case, $(\hbar\omega-\hbar\omega_0)^{3/2} \exp{[(\hbar\omega_0 -\hbar\omega)]/k_b T}$ when ($\omega >  \omega_0$) and a small Urbach type tail (for $\omega < \omega_0$) to account for low energy states extending into the band gap.  These first of these two $\Gamma$ factors is essential for quantitative agreement along the high energy side of the 0-0 PL band in $C_\beta$-type PL spectra (both at 18 K and higher temperatures).  $\Gamma$  also includes convolution with a narrow Gaussian representing instrumental resolution and residual disorder.  The only essentials needed in the following text are the HR parameters, the $\omega_k$'s (for use after the t.b.~calculations) and the displacement parameter, $D_{\mbox{\scriptsize rel}}=\sum_{i=1}^{9} {\hbar\omega_i S_i} $.  $D_{\mbox{\scriptsize rel}}$ measures the relative difference between the ground and excited state coordinates.
\begin{table}
\caption{Eq.~1 parameters from fits to PL data in Fig.\ 1.}
\begin{tabular}{ll|ll|ll|ll} \hline\hline
\multicolumn{2}{c}{Data set $\Rightarrow$} & \multicolumn{2}{c}{$C_\beta$ type}  
 &  \multicolumn{2}{c}{$C_{\gamma}$ type}   &  \multicolumn{2}{c}{$C_\alpha$ type}      \\ 
\multicolumn{8}{c}{~$\hbar \omega_0$/D$_{\mbox{\scriptsize rel}}$ (eV) ~~~2.806/0.12~~2.839/0.11~2.922/0.21}  \\ \hline 
\multicolumn{2}{c}{$\hbar \omega_i$ (eV)}    &  \multicolumn{6}{c}{HR factors ($i=$1 to 9)} \\ \hline
 1:~0.0075   & 6: 0.159  &   ~0.75 &  0.17~  &   ~0.98 & 0.13~ &  ~2.17 & 0.21~ \\ 
 2:~$\sim$0.07 & 7: 0.167  &   ~0.02  & 0.04~  &   ~0.03 & 0.08~ &  ~0.03 & 0.17~ \\
 3:~$\sim$0.09 & 8: 0.176  &   ~0.02  & 0.05~  &   ~0.03 & 0.05~ & ~0.03 &  0.05~ \\
 4:~$\sim$0.11 & 9: 0.199  &   ~0.02  & 0.30~  &   ~0.03 & 0.19~ &  ~0.03 & 0.46~\\
 5: 0.140   &         &   ~0.11  &  &  ~0.06 &  & ~0.14 \\ \hline \hline
\end{tabular} 
\end{table}

For structural modeling we used a combinatorial method approach in which a large number of possible single chain conformers were examined using empirically determined force-field parameters (MM3*\cite{maestro2}) at zero-temperature in the gas-phase.  Recent work in polysilanes\cite{psil:pdhsmodel:mjw03mm} and a related PF\cite{pfo:pf26:mm04mjw} suggest that this approach, though limited in scope, reveals some basic attributes of the local intrachain structure.  In this case over 700 distinct starting F8 decamers were constructed.  The fluorene backbone consisted of a repeating dyad (variants of a 2/1 helix as suggested in Ref.~\onlinecite{PFO:processXRD:mm99Bradley}) in which the starting torsion angles were alternately set to $\pm160 ^\circ$ respectively.  The three core alkyl side chain carbon-carbon dihedral angles (see Fig.\ 1) were then assigned to all possible combinations of the three conventional alkyl low energy conformers (i.e., {\it anti}, {\it gauche} and {\it gauche}$'$) within each alkyl chain.   Every fluorene monomer received the identical side chain conformer.   Additional details appear in Ref.~\onlinecite{pfo:pf26:mm04mjw}.  More extensive modeling (to identify viable dyad and larger conformational repeats) and molecular simulations are needed to fully validate this approach.   
\begin{figure}  \begin{center}
\includegraphics[width=3.1in]{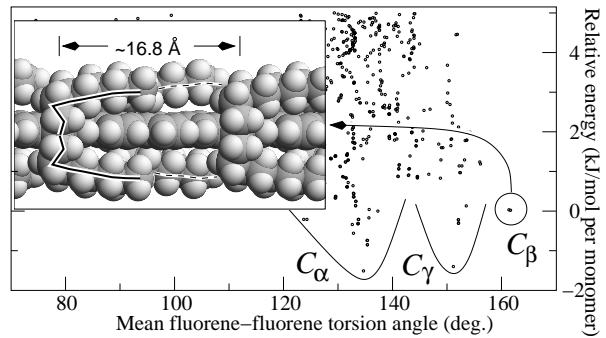} \\
\end{center}
\caption{ Low energy PF8 conformers (referenced to the $C_\beta$ isomer). Each symbol corresponds to a single, unique tested conformer.  $C_\alpha$, $C_\beta$ and $C_{\gamma}$ correspond to the proposed conformer families. Inset: Model of $C_\beta$ isomer highlighting side~chains.}   \end{figure}
Figure 2 displays the lowest zero-temperature energies of the trial starting conformers in terms of the average backbone torsion angle $|\phi|$.  At torsion angles above 120$^\circ$, there is some indication of  three distinct classes of low energy chain structures  near $|\phi|=$135$^\circ$, 150$^\circ$ and 160$^\circ$  (or $C_\alpha$, $C{_\gamma}$ and $C_\beta$ respectively).  Only a single unique isomer represents the C$_\beta$ type conformer.  The Fig.~2 inset depicts the resulting compact molecular construction.  The alkyl chains adopt a {\it trasoid} (i.e., nearly {\it anti}), {\it gauche}, {\it gauche} arrangement of the three innermost carbon-carbon torsion angles ($\tau_1$, $\tau_2$ and $\tau_3$ in the Fig.~1 inset).  Van der Waals interactions between neighboring alkyl side chains and the fluorene backbone work to stabilize this structure.  This conformer appears to be unstable if the alkyl chain length is two CH$_2$ units smaller and similar analysis of poly(di-2-ethyl-$n$-hexylfluorene)\cite{pfo:pf26:mm04mjw} gives only monomodal distributions of helices.   We therefore hypothesize the presence of three low energy classes of chain structures with differing average chain conformations and that $C_\beta$ segments are well ordered.

This suggestive finding is further tested by employing a conventional t.b.\ Hamiltonian\cite{Th:tightbind:jcp87silbey,psil:excite:hochstasser92jcp,Th:tightbinding:prl03knoester}  \begin{equation} H = \sum^N_{n=1}E_n |n\rangle\langle n| + \sum_{m=n+1}J_{mn} \cos(\phi_{mn}) |m\rangle\langle n| \end{equation} where $E_n$ and $J_{mn}$ are on-site excitation energy and nearest-neighbor intersite transfer energies respectively.  In this analysis we assume {\em no} on-site disorder and each of the two coefficients ($E_n$ and $J_{mn}$) are given a single unique value.  The second term has been modified by a cosine function to mimic the nature of $\pi$-conjugation\cite{Th:costheta:jcp96silbey} as the backbone torsion angle $\phi_{mn}$  varies. After diagonalization of the matrix the density of states and oscillator strengths were calculated in standard fashion to give an absorption lineshape $
A(\omega) = \langle \sum_j \mu_j^2 \delta(\omega - \omega_j) \rangle $
where $\mu_j^2$ is the transition dipole moment of the j$^{th}$ eigenstate and
averaged over the entire ensemble.  This function was further convoluted with the appropriate FC expression for absorption\cite{OPV:subband:jcp02Gierschner} using HR parameters identified
with the conformer dominating the respective PL (as in Table I) and a narrow Gaussian representing the spectrometer resolution plus residual disorder.  

Calculations of the PL employed $A(\omega)$ directly.  To model emission from excitons in thermodynamic equilibrium at the bottom of the band $A(\omega)$ was multiplied by a Boltzmann factor, $\exp{(\hbar\omega_0-\hbar\omega)/k_b T}$ ($\omega \ge \omega_0 $), next convoluted with the composite resolution/residual disorder function and then expressed using a conventional FC formula.   Matching the exact energies required a small Stokes shift defined as $\Delta$.  Clearly this simple approach cannot address the myriad of processes present in real PFs.  There is no accounting for either time-dependence or energy migration.  In addition there is nothing that reflects an exciton binding energy or interchain excitations.  Only singlet exciton emission is modeled.

In all these calculations each phenylene ring (i.e., $\frac{1}{2}$ of a fluorene unit) was deemed the basic structural unit so that every other off-site term was set to $-J_{mn}$ because of the chemically imposed planarization of the fluorene unit.  A single chain was truncated to 100 monomer units (or 200 sites) and this qualitatively reflects the experimental length of the PF8.  To represent  conformational isomerism we specified a different average fluorene-fluorene torsion angle, $|\phi|$, for each family and to this we typically added a Gaussian distribution with a standard deviation $\sigma$.  In most cases only two of these three claimed families were chosen by specifying $x$ and $1-x$ relative phase fractions assuming the minority phase segment length was dictated by Poisson statistics.  In this way features of both segmental and worm-like disorder\cite{psil:excite:hochstasser92jcp} are included. Over 4000 single chain trials were averaged for adequate statistics.  $A(\omega)$ was sensitive to even small parameter changes and thus many successive iterations were needed.

Figure 3 displays direct comparisons of both PA and PL in four representative data sets (three are from Fig.~1) with those calculated. Table II lists the associated parameters.  These parameters best approximate the leading edge of the PA; the region that is most critical for simultaneously assessing both PA and PL.  In particular the anomalous lineshape of the $C_\beta$ 0-0  is accurately reflected in this calculation.  The leading edge of the PA has an extended region over which a near power law rise occurs and it is this feature that gives the atypical lineshape on the high energy side of the 0-0 peak and, in regards to peak position and width, the systematic thermal progression in Ref.~\onlinecite{pfo:pf8band:prb03mjw}.  Replacing the Poisson distribution of the minority isomer segmental length with a Gaussian worsens these fits.   
This attribute is washed out in the $C_\alpha$ and $C_{\gamma}$ dominated PL spectra because $\sigma$ and the residual disorder (see text on $\Gamma$) are many times larger. 
\begin{figure} 
\begin{center}
\includegraphics[width=3.15in]{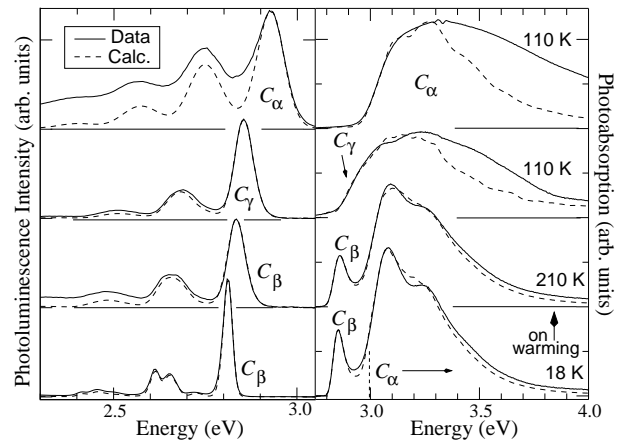}
\end{center}
\caption {PA and PL data in comparison with those obtained from the Eq.~2 based tight-binding calculations.}
\end{figure}

An important factor for estimating the relative phase fraction is the energy dependence of the oscillator strength.  Spectra similar to the top PA in Fig.~1 are assigned up to a 25\% $C_\beta$ concentration based on lifetime measurements\cite{PF:morph:prb04kohler}. If the actual t.b.~results are directly employed then the actual fraction drops to under 10\%.  This ``phase'', as suggested by a kinetic analysis\cite{pfo:pf8band:prb03mjw}, appears to be isolated one-dimensional threads of enhanced structural order embedded in sea of more disordered conformers.  Moreover these t.b.~calculations obtain relatively short segmental lengths for the emitting chromophore unit and, even in the locally well-ordered $C_\beta$ isomer, averages only five monomers.  
\begin{table}
\caption{Tight-binding model parameters.}
\begin{tabular}{r |cc c c c| c c c| c}
\hline\hline
\multicolumn{1}{c|}T & \multicolumn{5}{c|}{Minority conformer}   & \multicolumn{3}{c|}{$C_\alpha$-type region}   & $\Delta$ shift \\
(K) & \multicolumn{1}{c}{type}  &  conc.       & $\langle n \rangle $  & $|\phi|$ & $\sigma$    & $\langle n_\alpha\rangle$ & $|\phi_\alpha|$ & $\sigma_\alpha$   & (meV)\\
\hline
18  & $C_\beta$ & 9.5\%   & 5.0   & 165$^\circ$ & 3$^\circ$  & 32 & 136.5$^\circ$ & 13$^\circ$  & 10\\
210 & $C_\beta$                    & 9.3~~   & 5.1   & 165~ & ~6~~  & 33  & 135 &15~    & 24\\
110 & $C_\beta$                      & 0.7~~   & 3.0   & 172~ & ~3~~  & 55  & 129 & 31$^\dagger$    &~9\\
110 & $C_{\gamma}^\ddagger$ & 9.1~~  & 6.2  & 155~ & 15~~ & 29 & 127 &31$^\dagger$  & 11\\
\hline \hline
\multicolumn{10}{l}{$^\dagger$Assumes a uniform distribution of torsion angles.} \\
\multicolumn{10}{l}{$\ddagger$ Additionally incorporates 0.5\% of $C_\beta$ conformers.}\\
\end{tabular}
\end{table}

All calculated $|\phi|$ and $\sigma$ parameters adopt physically realistic values and, in regards to $|\phi|$, match the nominal oligomer modeling values.   Matching the temperature dependence of the $C_\beta$ dominated sample required few parameter changes except for $\sigma$ (which doubles at 210 K).    Increasing $\sigma$, because of the oscillator strength weighting effects, enhances the leading edge of the PA. The Boltzmann weighting factor provides for an immediate fit of the PL data although $\Delta$, a Stokes shift, goes from 10 to 24 meV.  This may reflect the nature of exciton energy migration.  The t.b.\ results yield $E_n$ and $J_{m,m}$ values of 5.45 and 1.35 eV.  These give a $\pi$-$\pi^*$ threshold energy of 2.75 eV and a band width of 2.70 eV.  This threshold approximates  those of the planarized ladder type polymers\cite{PFO:ladderline:prl03lupton}.
 
There are additional factors that need to be emphasized. Mimicking the slow rise on the leading edge of the PA in samples dominated by  $C_\alpha$ or $C_{\gamma}$ type emission required very small but finite fractions of $C_\beta$.    Modeling of the broad PA at higher energy in these two cases is imperfect. The quenched ($C_\alpha$ type) sample includes a uniform distribution of torsion angles between 98$^\circ$ and 160$^\circ$.  More complex distributions are necessary to address the failure at high energies.   Other shortcomings are present.  All FC progressions for the PAs assume fixed HR coefficients regardless of $\hbar\omega$.   The implicit heterogeneity of these PF8 models should produce systematic variations in these coefficients.  Moreover the PA HR coefficients are unlikely to be identical to those for PL. 

We also resolve an order dependent systematic failure of the FC model.  As the number of vibronic overtones rises the PL fit becomes progressively worse.  This even occurs on the low energy side of the 0-0 peak.  This effect may, in part, be due to the limited number of the $\omega_k$'s used.  However the 0-4 FC vibronic sub-band structure (near 2.1 eV) can still be resolved experimentally (see Fig.~1\cite{newresolution}) and this is many times that given by Eq.~1.  Given the simplistic assumptions used to specify the HR parameters this failing is not surprising\cite{Th:FCmulti:jcp03suzuki}.

One unexpected result relates to the $D_{\mbox{\scriptsize rel}}$ values listed in Table I.  Because the equilibrium excited state geometry approaches planarity\cite{Th:flatfluorene:mm04zhang01} we naively expect emission from the most planar ground state conformations (i.e., $C_\beta$-type) to yield the smallest displacement parameter.  We actually find that the smallest $D_{\mbox{\scriptsize rel}}$ occurs when the PF8 PL is dominated by the claimed $C_{\gamma}$-type family.   Quantum chemical calculations and other model studies are still needed to establish the veracity of this result.

As a final comment we note that this discrete three family model is overly abrupt and that one may expect more gradual crossovers in terms of local structure.  The modeling procedure itself has merit for identifying prospective chain conformations including the side chains and, with additional effort, conformational disorder as well.   Successfully deciphering the single chain properties is one very important step for developing a full microscopic understanding of structure and photophysics. 

NSF DMR-0350383 grant support is gratefully acknowledged.  

\bibliographystyle{apsrev} 
\bibliography{/home/winokur/bib/ppv,/home/winokur/bib/lcp,/home/winokur/bib/led,/home/winokur/bib/theory,/home/winokur/bib/oops,/home/winokur/bib/psil,/home/winokur/bib/pfo} 
\end{document}